\documentstyle[prl,aps,epsfig]{revtex}

\begin{document}

\draft
\flushbottom
\twocolumn[\hsize\textwidth\columnwidth\hsize
\csname@twocolumnfalse\endcsname
\title {
Nanoscale Phase Coexistence and Percolative Quantum Transport
}

\author{Sanjeev Kumar and Pinaki Majumdar}

\address{ Harish-Chandra  Research Institute,\\
 Chhatnag Road, Jhusi, Allahabad 211 019, India }

\date{Nov 15,  2003}

\maketitle
\tightenlines
\widetext
\advance\leftskip by 57pt
\advance\rightskip by 57pt

\begin{abstract}

We study the nanoscale phase coexistence of ferromagnetic 
metallic (FMM) and antiferromagnetic insulating (AFI)  regions by including 
the effect of AF superexchange and weak disorder in the  double exchange 
model.  We use a new  Monte Carlo technique, mapping on 
the disordered spin-fermion problem to an effective short range spin model, 
with self-consistently computed exchange constants.  We recover `cluster 
coexistence' as seen earlier in exact simulation of small systems. The much 
larger sizes, $\sim 32 \times 32$, accessible with our technique, allows 
us to study the cluster pattern for varying electron density, disorder,
and temperature.
We track the magnetic structure, obtain the density of states, with its 
`pseudogap' features, and, for the first time, provide a fully microscopic  
estimate of the resistivity in a phase coexistence regime, comparing it with
the `percolation' scenario.

\

\

\end{abstract}

]

\narrowtext
\tightenlines

The issue of multiphase coexistence in transition metal oxides 
has been brought to the fore by a set of remarkable recent experiments
on the manganites
\cite{math-litt-ssc,dag-rev}.
 These experiments probe the atomic scale magnetic 
correlations \cite{magn-image}, structural features  
\cite{latt-image}, or `conducting' properties, 
{\it i.e}
tunneling density of states
\cite{cond-image}, 
 through local spectroscopy.
The bulk thermodynamic and transport properties of these systems,
including the  `colossal magnetoresistance' (CMR) seem
to have a correlation with the physics of electronic phase coexistence
as visualised in these nanoscale experiments. 

Establishing a {\it first principles theoretical 
connection} between cluster
coexistence and `anomalous' transport, however, has 
been difficult. Only real space Monte Carlo (MC) techniques
\cite{dag-clust-mc}
allow access to the cluster phase, 
and  accessible sizes, $\sim 8 \times 8$ in two dimension (2d),
are too small to study transport. 
The  approach in current use
\cite{dag-res} is phenomenological, 
using classical resistor
networks to  model the  data within a `percolation
scenario'. 
While `percolation' surely plays  an important role in transport,
{\it standard } percolation theory \cite{stauffer} 
is inadequate
\cite{dag-perc}
to explain the data.
More fundamentally, such an approach does not explain the 
temperature and field dependent  
resistance of the `building blocks', or correlations and hysteresis
in the cluster distribution, or the dependence of bulk resistance
on `spin overlap' between clusters.

We address these questions within a `spin-fermion' model
in this paper, employing a new MC technique  \cite{sk-pm-mc}
that allows access to
system sizes $\sim 32 \times 32$ in 2d.
This can probe the regime 
$L_S \gg {\bar L_c} \gg a_0$, where $L_S$, ${\bar L_c}$ and $a_0$ are
system size, typical cluster size, and lattice spacing, respectively.
We study the coexistence regime, obtain the typical 
cluster size, and calculate the spectral density and 
conductivity in the mixed phase. 
To our knowledge this is the first microscopic calculation 
to clarify the connection between 
nanoscale phase coexistence and transport 
in a fully quantum mechanical itinerant
electron system.

The coexistence of two phases with distinct electronic,
magnetic, and possibly structural, properties is 
best conceived at $T=0$ in a {\it clean system}.
For a specified chemical potential, $\mu$, the ground state
configuration of the spin and lattice variables, $\{ X_i \}$
say,
assumed classical, 
is determined by $\delta {\cal E}/{\delta  X_i} =0$, where
${\cal E}\{ X_i ; \mu\}$ is the energy of the system in the
$\{ X_i \}$ background. 
The minimum, ${\cal E}_{min}(\mu)$, 
usually occurs for a unique $\{ X_i \}$ 
at each $\mu$, and
in this  background the electron density $n(\mu)$ is also
unique. However, in the presence of competing interactions,
two {\it distinct} $\{ X_i \}$ 
configurations could be degenerate
minima  of ${\cal E}$ at some $\mu$. 
The corresponding 
$\mu = \mu_c$ marks a {\it first order
phase boundary}, and the two `endpoint' densities, $n_1$ and 
$n_2$ bracket a region of coexistence. 
There is no homogeneous phase with density between $n_1$ and $n_2$.
In this regime the system breaks up
into two macroscopic domains, with density $n_1$ and $n_2$.

It was pointed out by Dagotto and coworkers \cite{dag-clust-mc},
and known earlier in the context of 
classical spin models \cite{imry-ma},
that disorder is a singular perturbation in these  systems.
Even weak disorder breaks up the `macroscopic' domains into
{\it interspersed locally ordered  clusters of the 
two phases.} This nano/mesoscopically inhomogeneous 
strongly correlated
phase can be studied only within a real space framework.

In this paper 
we study the key qualitative issues of phase coexistence in  itinerant
fermion models by considering the competing effects of double exchange (DE)
and superexchange (SE)
in the presence of weak
disorder. We consider   
$H = H_{el} + H_{AF}$ in 2d, with: 
\begin{equation}
H_{el} =
\sum_{\langle ij \rangle \sigma} t_{ij} 
c^{\dagger}_{i \sigma} c^{~}_{j \sigma}
+  \sum_{i } (\epsilon_i - \mu) n_{i}  
- J_H\sum_i {\bf S}_i {\bf .} {\vec \sigma}_i 
\end{equation}
and 
$  H_{AF} = J_S\sum_{\langle ij \rangle} {\bf S}_i.{\bf S}_j $.
The hopping $t_{ij}=-t$ for nearest neighbours, and $\epsilon_i$ is
uniformly distributed
between $\pm \Delta/2$. 
We assume $J_H/t \rightarrow \infty$.
The parameters in the problem are  $\Delta/t$,
$J_S/t$ and density $n$ (or chemical potential $\mu$). 
We use $\vert {\bf S}_i \vert =1$.
For $J_H/t \rightarrow \infty$, 
the Hamiltonian, in the projected basis \cite{sk-pm-mc},
assumes a simpler form:
$ H_{el} 
= -t\sum_{\langle ij \rangle} f_{ij}
(~e^{i \Phi_{ij}}  \gamma^{\dagger}_i  \gamma_j + 
h.c~) + 
\sum_i (\epsilon_i - \mu) n_i $.
The hopping amplitude, $g_{ij} = f_{ij} e^{i\Phi_{ij}}$, 
between locally aligned states,
can be written in terms of the polar angle $(\theta_i)$ and
azimuthal angle $(\phi_i)$ of the spin ${\bf S}_i$ 
as,
$  cos{\theta_i \over 2} cos{\theta_j \over 2}$ 
$+
sin{\theta_i \over 2} sin{\theta_j \over 2}
e^{-i~(\phi_i - \phi_j)}$.
The `magnitude'  of the overlap is 
$f_{ij} = \sqrt{( 1 + {\bf S}_i.{\bf S}_j)/2 }$,
while the phase is specified by 
$tan{\Phi_{ij}} = Im(g_{ij})/Re(g_{ij})$.

To make progress we need the `effective Hamiltonian' controlling
the Boltzmann weight for 
the spins.
Formally, this is:
$ H_{eff} \{ {\bf S}_i \} = 
 -(1/\beta) logTr e^{-{\beta}( H_{el} + H_{AF})}$.
The `exact' MC directly 
generates {\it equilibrium spin configurations} of 
$H_{eff}$ through
iterative diagonalisation \cite{dag-clust-mc}. 
We make the  approximation:  
$ H_{eff} =
-\sum_{\langle ij \rangle} D_{ij} f_{ij}  
+ J_S\sum_{\langle ij \rangle} {\bf S}_i.{\bf S}_j$,
with  $D_{ij}$ 
being  determined 
self consistently \cite{sk-pm-mc} as 
the average of   
${\hat \Gamma}_{ij}
= (e^{i \Phi_{ij}}  \gamma^{\dagger}_i  \gamma_j + h.c)$
over the assumed equilibrium distribution.
This approximation has been extensively benchmarked for
the clean DE model \cite{sk-pm-mc} and we will put up similar 
comparisons for the present problem in the near future 
\cite{sk-pm-meso-unpub}.

Although the $D_{ij}$ enter as `nearest neighbour'
exchange, they  arise from a solution of the
{\it full quantum statistical  problem in the disordered 
finite temperature system.} 
In the presence of competing interactions and quenched disorder,
this leads to a set of 
strongly inhomogeneous, spatially correlated, and temperature
dependent `exchange' $D_{ij}$.

At consistency, fermionic properties
are computed and averaged over  equilibrium spin configurations.
We work 
\begin{center}

\vspace{.2cm}

\begin{figure}
\hspace{-.3cm}
\epsfxsize=7.5cm \epsfysize=6.60cm \epsfbox{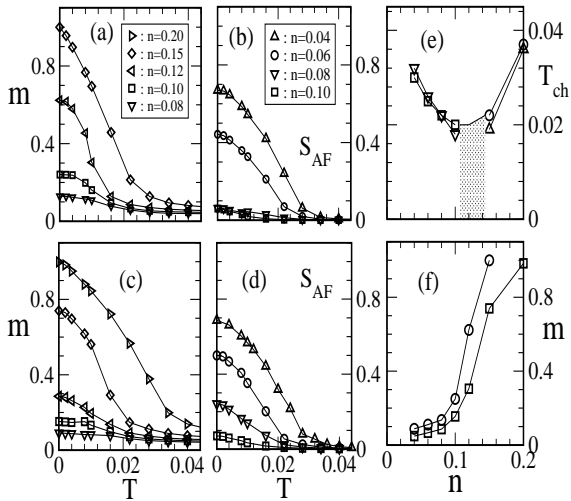}

\vspace{.1cm}

\caption{Magnetisation, $m(T)$, and structure factor, $S_{\bf q}(T)$ at
${\bf q} = \{ \pi, \pi\}$. Panel $(a)-(b)$: data for $\Delta=0.2$, panel
$(c) - (d)$: data for $\Delta=1.0$. 
The legends are common to panels $(a)-(d)$.
Panel $(e)$: The characteristic temperature (see text), 
$T_{ch}(n, \Delta)$:
triangle down: AF at $\Delta=1.0$, square: AF at $\Delta=0.2$, 
triangle up: FM at $ \Delta=1.0$, circles: FM at $\Delta=0.2$.
Panel $(f)$: $m(T=0)$ with varying $n$, circles: $\Delta=0.2$,
squares: $\Delta=1.0$.
System size:  $24 \times 24$.
}
\end{figure}
\end{center}
\begin{center}
\begin{figure}

\epsfxsize=6.0cm \epsfysize=4.0cm \epsfbox{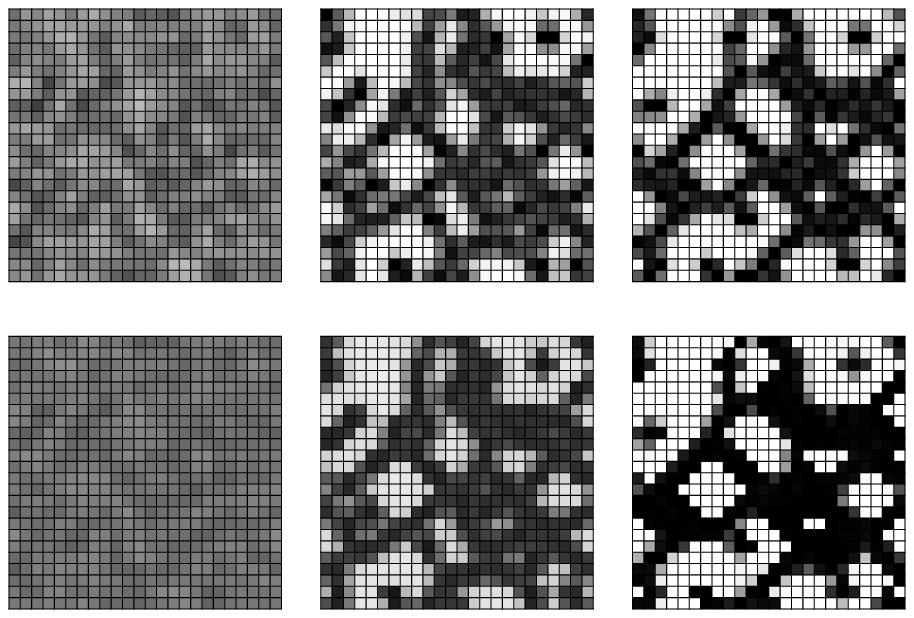}
\vspace{-.0cm}
\epsfxsize=6.0cm \epsfysize=4.0cm \epsfbox{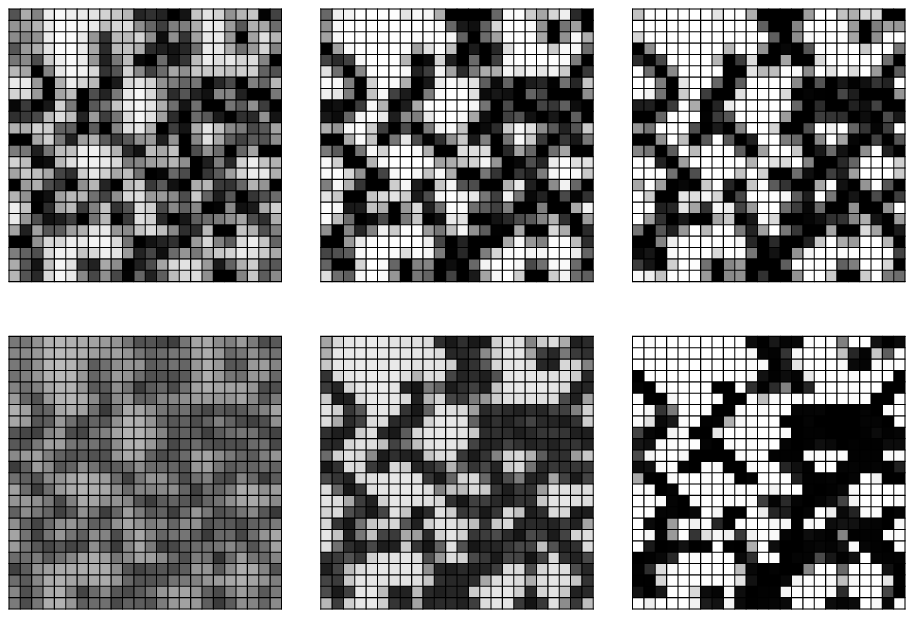}

\vspace{.2cm}

\caption{
Cluster pattern: Rows 1-2 are for 
$\Delta = 0.2$, row 3-4 are for $\Delta=1.0$.
Mean density $n=0.1$ in both cases.
Upper row in each set shows the thermally averaged density profile,
$ n_{\bf r}$, the lower row shows the nearest neighbour spin correlation:
$f_2^s = \langle {\bf S}({\bf r}).{\bf S}({\bf r} + 
{\bf \delta}) \rangle$. 
Left to right along each row,  the $T$  variation is 
$T/T_c  \sim 2.0, 1.0,  0.0$.
For $n_{\bf r}$ dark regions are high density. For $f_2^s$
dark regions are FM, white regions AF.
System size $32 \times 32$, data for a specific $\{\epsilon_i \}$
in each case.
}

\end{figure}
\end{center}
at specified mean density, fixing $\mu$ through iteration.
Transport properties 
are computed 
using the Kubo formula \cite{sk-pm-mc}, 
employing sizes 
$\sim 24 \times 24$ to $32 \times 32 $, 
and 
averaged typically over $30$
realisations of disorder, with averaging over $\sim 50$ equilibrium spin
configurations at each $T$ for each realisation of $\{\epsilon_i \}$.
Since the d.c conductivity is not directly accessible in a finite
system,  we 
compute the finite frequency  conductivity at the scale
of mean level spacing, $\omega_L \propto  8t/L^2$, with $\omega_L 
= 0.04t$ at $L=32$.
The conductivity results are in units of $(\pi e^2)/{\hbar}$.
We have checked the adequacy of disorder average and stability 
with respect to system size variation \cite{sk-pm-meso-unpub}.

Our principal results include the direct 
visual evidence of cluster 
coexistence (Fig.2),  the disorder dependence of 
spectral density (Fig.3) and the 
resistivity (Fig.4) in the coexistence regime.

At low electron density, the competetion in the DE$+$SE model 
is between a $\{ \pi , \pi \}$  AF phase and a ferromagnet (FM). 
We set $J_S = 0.05$ and scanned in $\mu$ to
locate the $\mu_c$ for the first order boundary. The density
changes from $n=0$ to $n\sim 0.20$ at the discontinuity. These results
were obtained within the $H_{eff}$ scheme and cross checked with
exact MC.
Moderate disorder smooths out the discontinuity in
$n(\mu)$ converting it to a sharp crossover
\cite{sk-pm-meso-unpub}. 
In this paper we fix  $J_S = 0.05$ and probe the coexistence
regime, $n \sim 0-0.2$, with varying electron density, temperature $(T)$,
and $\Delta = 0.2$ and $\Delta=1.0$.
All the results are obtained by cooling the system from the 
high  $T$ 
\begin{center}

\begin{figure}
\epsfxsize=8cm \epsfysize=5.50cm \epsfbox{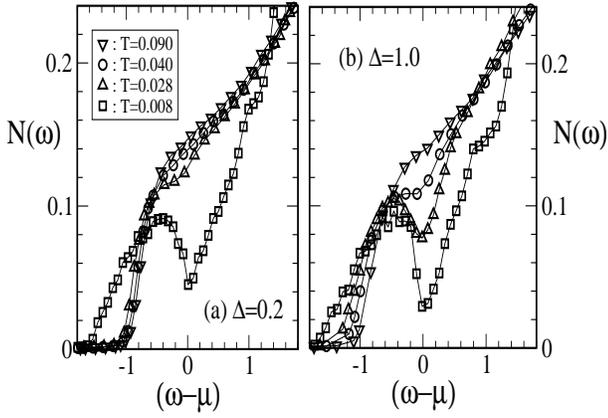}

\vspace{.2cm}

\caption{Low energy density of states, 
at $n=0.10$.
Panel $(a)$: data at  $\Delta = 0.2$, panel $(b)$: data at
 $\Delta = 1.0$. The DOS is plotted with respect to $\omega - \mu(T)$,
incorporating the $T$ dependent shift in $\mu$. Lorentzian broadening of
$\delta$ functions to $0.03$. Data averaged over 30 copies.}
\end{figure}


\end{center}
paramagnetic, approximately homogeneous, phase.

Fig.1 shows the magnetisation, $m(T)$, and the AF peak in the 
magnetic structure factor $S_{\bf q}$
at ${\bf q} = \{ \pi, \pi \}$ to illustrate the evolution 
from the AF to the 
FM 
phase with increasing $n$.
Panel $(a)-(b)$ are at $\Delta=0.2$ and panel $(c)-(d)$ at
$\Delta=1.0$. 
The `extremal' densities are either strongly AF or FM,
while for $n \sim 0.08-0.12$ {\it both   
FM and AF
reflections have finite weight}. Panel $(e)$ tracks the
`characteristic temperature', $T_{ch}(n, \Delta)$, 
identified from the maximum in ${\partial^2O}/{\partial T^2}$ where
$O$ is the appropriate order parameter (of the FM or AF phase).
In the shaded region, $0.10 
\lesssim  n \lesssim  0.14$,
it is difficult to  resolve the $T_{ch}$ accurately.
Panel $(f)$ shows the change in `saturation magnetisation'
$m(T=0)$ with increasing $n$ and changing disorder. 
The small moment regime, $m \lesssim 0.1$, for $n \lesssim 0.1$
is a `ferro-insulator' phase, as we will discover from the transport
data. In this regime  
the  moments in different clusters are only weakly correlated.

Fig.2 provides direct visual evidence of `clustering',
the top two panels showing thermally averaged 
local density and nearest neighbour
spin correlation for  $T/T_c \sim 2.0, 1.0, 0.0$ (left to right),
at $\Delta=0.2$. The
lower panels show  data at $\Delta=1.0$.  We will
quantify the `typical scale' of clusters later, but from  
the density distribution it is obvious that the pattern for
$\Delta=1.0$ is more fragmented than for $\Delta=0.2$.
The 
`density contrast' reduces with increasing $T$, as the spins in the
AF regions fluctuate out of  antiparallel alignment, and the
carriers can partially explore these regions. 
 The spin correlations
also weaken, with FM (black) and AF (white) regions giving way
to an uncorrelated (grey) background.

Fig.3 shows the low energy density of states (DOS) for $n=0.1$.
Even at weak disorder, panel $(a)$,
there is a `pseudogap' in the system 
averaged DOS, at the lowest temperature, $T \sim 0.01$. 
However, the pseudogap rapidly fills up with increasing $T$,
and the DOS tends towards 
\begin{center}

\begin{figure}
\epsfxsize=8.0cm \epsfysize=5.6cm \epsfbox{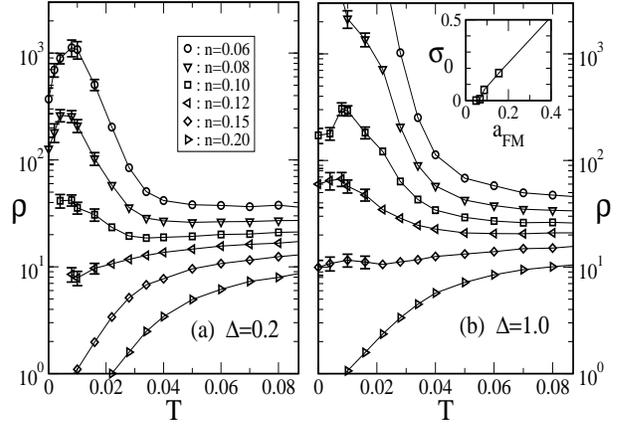}

\vspace{.2cm}

\caption{Resistivity, $\rho(T)$, with varying $n$. Panel $(a)$: $\Delta =0.2$,
panel $(b)$: $\Delta=1.0$.  
Results obtained on `cooling'. System size $24 \times 24$, data obtained by 
inverting the `mean conductivity'. Average over
$20-50$ copies of $\{ \epsilon_i \}$ (error bars comparable to symbol 
size, unless otherwise indicated).
 Results on size $32 \times 32$ are 
similar.
Inset, panel $(b)$: Normalised  $T=0$ conductivity, at $\Delta=1$,
  -vs- 
FMM surface area.
}
\end{figure}


\end{center}
the universal profile of the spin disordered
2d DE model. 
At stronger disorder, panel $(b)$,
 the `dip' in the DOS at $\mu $ is
deeper. Due to stronger pinning, the clusters, and the 
gap feature, survives to higher $T$, 
and only for $T \gg T_{ch}$ tends to the asymptotic form.
While this data focuses on the system averaged DOS, which can be
probed by photoemission (PES), tunneling spectroscopy 
would track 
the {\it local } DOS, which shows \cite{sk-pm-meso-unpub}
a clean `gap'
in the AF regions, and finite  DOS 
in the (larger) metallic 
clusters. 

Fig.4 shows  resistivity, $\rho(T)$, with varying
$n$ and disorder. The comparison of panel $(a)$, $\Delta=0.2$,
and panel $(b)$, $\Delta=1.0$,
indicates that $\rho(T)$ for stronger disorder is systematically
larger.
The {\it trends}, however, are similar in the two cases and
allows a tentative classification of the `global' aspects of
transport.
At both $\Delta=0.2$ and $\Delta=1.0$ there is a critical density,
$n_c(\Delta)$, below which $\rho(T=0)$ diverges, indicating the 
absence of any connected `conducting path'.
The inset to panel $(b)$ shows the trend in the $T=0$ conductivity with
increasing  FM cluster area, $a_{FM}$. 

We suggest the following qualitative picture of transport
in the coexistence regime based on the data in Fig.4. 
$(a)$.~At low $T$, and for $n > n_c(\Delta)$, there is, by definition,
some
connected `metallic path' through the sample. It is reasonable to
assume that at least in this connected region the spins are aligned
due to DE. 
Since cluster dimension and
electron wavelength are comparable in our system 
a large fraction of the resistance arises from the non trivial geometry
of the current path. 
This scattering is  quantum mechanical, and the low $T$ metallic 
phase corresponds to a 
{\it quantum percolative regime} \cite{avish}, for 
roughly $0.15 \gtrsim n \gtrsim
0.08$ at $\Delta=1.0$.  
$(b)$.~With increasing $T$, the resistance of the conducting network
increases due to DE spin fluctuations and,  till  the network
is  disrupted, 
there 
\begin{center}
\begin{figure}

\epsfxsize=8.00cm \epsfysize=5.50cm \epsfbox{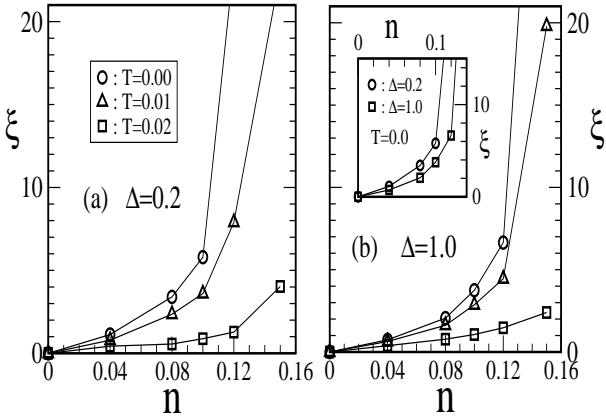}

\vspace{.4cm}

\caption{Typical FMM cluster size:   $(a)$~$\Delta=0.2$ and varying $T$,
$(b)$~$\Delta = 1.0$,  varying $T$. The inset to panel $(b)$~highlights
the reduction in cluster size with increasing disorder.
}

\end{figure}
\end{center}
is a regime $d\rho/dT >0$.
This `metallic' behaviour occurs 
despite the very large residual resistivity.
This is a regime of {\it weak magnetic scattering on
the conducting network}.
If $n \gg n_c$, {\it e.g},  $n=0.20$ in Fig.4.$(b)$,
so  that inhomogeneities are weak,
$\rho(T)$ will smoothly increase to the $T \gg T_{ch}$ asymptotic 
 value.
$(c)$.~With further increase in $T$, in the $n \gtrsim n_c$ regime,
the spin disorder can destroy
some of the `weak 
links', disrupting the conducting network and
leading to a sharp increase in $\rho(T)$ see, 
{\it e.g},  $n=0.1$ in Fig.4.$(b)$. This correlates well with reduction
in typical cluster size with increasing $T$, discussed in the next
paragraph.
Depending on $n$ and
the extent of disorder (and 
system size, in a simulation) there
could be a rapid rise 
or a `first order' 
metal-insulator transition (MIT).
This is  
{\it spin disorder induced breakup of
clusters driving a MIT.}
$(d)$.~ Beyond this `MIT'
the conduction is through the `insulating' regions, 
with isolated patches contributing to nominally 
activated transport, as visible in Fig.4.$(b)$
for $n \lesssim 0.12$.
For $T \gg T_{ch}$, as the structures disappear, Fig.2, and the system
becomes homogeneous, $\rho(T)$ is dominated by spin disorder
scattering. This is a {\it diffusive regime with saturated
spin disorder scattering}, visible for $T \gtrsim 0.08$ at all
densities. 
$(e)$.~For $n < n_c$, the regime of low density isolated clusters, 
$\rho(T)$ falls monotonically, and the response is typical of
{\it low density ferromagnetic polarons in a AF background.}
This occurs for $n \lesssim 0.08$ at $\Delta=1.0$, and at lower
$n$ at $\Delta=0.20$.

Finally, Fig.5 shows the typical size of FMM clusters, inferred from a
Lorentzian fit to the magnetic structure factor,  {\it i.e},
$S_{\bf q} \sim (q^2 + \xi^{-2})^{-1}$. The resulting correlation length
depends on $n$, $\Delta$ and $T$, decreasing with increasing disorder
and $T$, and increasing with increasing density. The main panels, $(a)$ 
and $(b)$, highlight the $n$ dependence at different $T$ and $\Delta$,
while the inset in Fig.5.(b) replots the same data to highlight the
dependence on disorder. The dependences, overall, are intuitive, and
strengthen the proposed transport scenario.

Our results on coexistence  share several generic features with 
the manganites \cite{magn-image,latt-image,cond-image},
but there are  key differences too.
Apart from the difference between 2d and 3d, these are
$(i)$~the important regime of coexistence in manganites is
between FMM  and charge ordered insulator (COI), 
and crucially involves the Jahn-Teller (JT)  phonons.  This 
requires enlarging our model. $(ii)$~It is supposed that the FMM
and COI regions are nominally of {\it equal density}, which is
why such domains can survive over $\mu $m scale. Our clusters are
`charged' due to the density difference between FMM and AFI.
In any real system Coulomb  effects  
\cite{kunyang} would have 
to be considered in such a case. We will discuss such effects
separately \cite{sk-pm-meso-unpub}, as well as the effect of an
applied magnetic field on transport.

In conclusion, we have presented microscopic results on
quantum transport across a regime of phase coexistence 
in an  itinerant fermion
model, and correlated it with thermal evolution of the 
spatial structures.
The method readily generalises to the two orbital model with
JT phonons, to be discussed in the near future.

We acknowledge use of the Beowulf cluster at H.R.I

{}

\end{document}